# Magneto-Dielectric Behavior in $La_{0.53}Ca_{0.47}MnO_3$

Suchita Pandey, Jitender Kumar, and A.M. Awasthi*
Thermodynamics Laboratory, UGC-DAE Consortium for Scientific Research,
University Campus, Khandwa Road, Indore- 452 001 (India)

## Abstract

We prospect magneto-dielectricity in $La_{0.53}Ca_{0.47}MnO_3$ across its paramagnetic (PMI) to ferromagnetic (FMM) isostructural transition at $T_C \sim 253$K, by magnetic ($M$), caloric ($W$), dielectric ($\varepsilon'$), magneto-resistive (MR), and magneto-capacitance (MC) investigations. Skew-broadened first-order transition character is confirmed via heating/cooling hystereses in $M(T)$ and $W(T)$, with superheating temperature $T^{**}$ almost next to $T_C$ and supercooling temperature $T^{*}$ exhibiting kinetics. Above $T_C$, linearly-related MC and MR reflect purely magneto-resistance effect. Near $T_C$, the high-frequency MC(5T) much exceeds the magneto-losses, and is uncorrelated with dc MR(5T) in the FM-ordered state. The intrinsic magneto-dielectricity manifest below $T_C$ and above ~kHz is traced to an intra-granular Maxwell-Wagner-type effect at the interface-region of PMI-FMM phase-coexistence.

* Corresponding Author e-mail: amawasthi@csr.res.in. Tel: +91 731 2463913.

## 1. Introduction

Intense research on magneto-electric (ME) multiferroics is driven by their applicability in the data-storage devices and spintronics. Multiferroicity of various origins in manganites has been reported recently in several systems; e.g., in hexagonal-$YMnO_3$ oxides [1-4] due to the geometric frustration, and in Perovskites $XMnO_3$ [5, 6] and $XMn_2O_5$ (X = Tb, Dy) [3, 7-8] due to the spiral magnetic ordering. Yet extremely rare is coexistent ferro -electric and -magnetic orders in single-phase compound. While $BiMnO_3$ does exhibit ferromagnetism ($T_C$ =100K) and ferroelectricity [9-10] ($T_C$ =450K), in $BiFeO_3$ a weak ferromagnetism (WFM) at room temperature is found [11], in concert with the coexistent ferroelectric ($T_C \sim$ 1103K) and antiferromagnetic ($T_N \sim$ 643K) phases. Thin films of $BiFeO_3$ reported as ferro-electromagnet make it the yet-only known 'room temperature' multiferroic [12]. Furthermore, over the past few years, several charge-ordering CMR manganites viz., $Pr_{0.6}Ca_{0.4}MnO_3$, $Nd_{0.5}Ca_{0.5}MnO_3$, $Pr_{0.7}Ca_{0.3}MnO_3$, and $La_{0.25}Nd_{0.25}Ca_{0.5}MnO_3$ [13], $Pr_{0.5}Ca_{0.5}Mn_{0.975}Al_{0.025}O_3$ [14], $La_{0.5}Ca_{0.5}MnO_3$ [15], $La_{0.67}Ca_{0.33}MnO_3$ [16], and $La_{0.5}Ca_{0.5-x}Sr_xMnO_3$ ($0.1 \leq x \leq 0.4$) [17] as magneto-dielectrics, have extended the multiferroics family [18]. Well-studied at different doping levels, $La_{1-x}Ca_xMnO_3$ (LCMO) shows charge ordering (CO) and magnetically-ordered (FM and/or AFM) ground states, depending upon the dopant concentration. In LCMO and similar systems (with different dopant-concentration, $x$) the magnetic ground state has been debated, due to varying relative proportions of ferro- and antiferro-magnetic phases [19].

Huang *et al.* [20] showed via neutron diffraction isostructural-changes in $La_{1-x}Ca_xMnO_3$ ($x$=0.47, 0.5, 0.53) spanning its AFM ($T_N$) and FM ($T_C$) transitions (minor ~14% AFM phase in $x$=0.47 specimen); the broad 170-260K window generally attributed to the competition between the two magnetic-orderings. The transition causing substantial but continuous increase (decrease) in $a$, $b$ ($c$) lattice parameter accompanies a weak Jahn-Teller distortion of the $MnO_6$ octahedra [20]. Resultant orbital ordering of the $Mn^{+3}$ and $Mn^{+4}$ (double-exchange interaction between $Mn^{+3}$-O-$Mn^{+4}$. Zigzag chains in $a$-$c$ plane) [21] is responsible for the ferromagnetic order at $T_C$. A rather broad $\{\sim O(T_C\text{-}T_N)\}$ range of isostructural-changes obscures a clear resolution of the two merged-up/overlapping transitions. Furthermore, a recent theoretical work on $La_{0.5}Ca_{0.5}MnO_3$ (LCMO50) by Giovannetti *et al*. [22] proposed that the combination of a peculiar charge-orbital ordering and a tendency to form spin-dimers breaks its inversion symmetry, and should result in a polar state with relatively strong magneto-dielectricity. Subsequently, a dielectric study [15] on LCMO50 reported a broad $\varepsilon'$-peak at 200K (*in between* close-by $T_C$ and $T_N$ [15], but *not exactly at either*), described as a combination of relaxor-ferroelectricity and long-range FE-order. We believe the anomalous feature reported (in the overlapping region of the two magnetic-transitions) is due to the coexistence of ferromagnetic and anti-ferromagnetic phases. Intriguingly, scanning thermo-grams of $La_{1-x}Ca_xMnO_3$ also reported [23] for half-doped composition (not for $La_{0.53}Ca_{0.47}MnO_3$), an anomalous absence of the FM-transition; attributed to a minimization in LCMO50 of the lattice-polarons' interaction (causing local lattice distortion). The latter was related to the offsetting-effects of hole-concentration and the ionic-radii-differences between $La^{3+}$, $Ca^{2+}$, $Mn^{3+}$, and $Mn^{4+}$. All these results mask a precise fixation of the transition at $T_C$; whether it is a continuous one generally taken [24], or a disorder-broadened first-order type [25], mandating its thermal and kinetic/hysteretic investigations.

To probe resolvable effects of the two magnetic phases on possible multiferroicity, we have chosen the composition $La_{0.53}Ca_{0.47}MnO_3$ (LCMO47) in which the FM-order is reported to occur at $T_C$ ~260K, followed by the CO-AFM transition at $T_N$ ~160K [20]. According to the phase diagram [19] LCMO47 also has adequately-separated coexistence-windows of para-ferromagnetic and ferro-antiferromagnetic phase-pairs, in contrast to the LCMO50, which has mainly one merged-up (FM-AFM) coexistence region. The FM transition focused here is more important & rare [9] regarding magneto-electricity, which we explore across $T_C$. The main issues addressed are; (*i*) to characterize the magneto-structural phase-transition, and (*ii*) to explore intrinsic magneto-dielectricity derived from coexistent para-ferromagnetic phases, unconnected to the finite magneto-resistance. We obtain unambiguous signatures in permittivity ($\varepsilon'$) and heat-flow ($W$) near $T_C$, also confirmed as the PMI-FMM transition by magnetization ($M$) and resistivity ($\rho$). Independent $M(T)$ and $W(T)$ warming/cooling data establish the first-order character of the FM transition. Our high-frequency $\varepsilon'_H(T)$ results attribute the genuine (dc $\rho_H(T)$-independent) magneto-dielectricity below $T_C$ to the PM-FM phase coexistence feature of the disorder-broadened ferromagnetic transition.

## 2. Experimental Details

Ceramic LCMO47 samples were prepared by the standard solid state reaction route with high-purity (99.99%) powders of $La_2O_3$, $CaCO_3$, and $MnO_2$. As most of the rare-earth oxides absorb moisture from the air, $La_2O_3$ was preheated at 1000 °C for 12h. All the powders were mixed in stoichiometric ratio, ground and calcined at, 1100, and 1250 °C for 24h, and then pressed into disk-shaped 2mm thick pellets of diameter 10mm. The pelletized samples were sintered at 1300 °C for 20h. XRD characterization of the samples was done with Bruker D8 advanced diffractometer using Cu-$K_\alpha$ ($\lambda$ = 1.54Å) radiation. It was found that the samples formed in single phase with no detectable secondary matter, and the diffraction pattern could be indexed purely with orthorhombic (*Pnma*) Perovskite-like crystal structure [26], with Rietveld-refined lattice parameters $a$ = 5.4294(3)Å, $b$ = 7.6580(4)Å, and $c$ = 5.4419(3)Å. Iodometric-titration confirmed stoichiometric-oxygen; its uncertainty (δ) found as ±0.0019 per formula unit [27]. Magnetization measurements were performed under 500 Oe field (heating-cooling at 1.2K/min) with the Quantum Design MPMS Vibrating Sample Magnetometer. Dielectric measurements were performed over 30Hz-1MHz frequency with 1V ac-excitation, using NOVO-CONTROL (Alpha-A) High Performance Frequency Analyzer. We used parallel-plate copper electrodes, with silver-paste-coated sample-surfaces. In-field (5T) dielectric characterization was done using an OXFORD Nano Science Integra System along with the NOVO-Control Analyzer. Heat flow thermograms in warming/cooling cycles and specific-heat were obtained over 150-300K using Modulated Differential Scanning Calorimeter (MDSC, TA-Instruments model TA-2910). Resistivity measurements were done with standard 4-probe method using home-made insert along with 8T Superconducting Magnet System (Oxford Instruments, UK).

## 3. Results and Discussion

*3.1 Magnetization and Specific Heat*

Figure 1a shows the *M-T* curve measured in 500 Oe field (1.2K/min warming), with clear PM to FM transition at ~253K. Figure 1b shows the heat-capacity obtained from our MDSC heat-flow (warming) data. Nearly coincident temperatures of the $C_p$-peak and the $M(T)$-inflexion point (taken at 1.2K/min warming) affirm little kinetic effects in the warming cycles (no phase-coexistence above $T_C$). As discussed later, our resistivity data too exhibit *purely* insulating behavior exactly down to 254K, and the dielectric peak anomaly is observed at 252K. These observations fix the ferromagnetic-structural $T_C$ = 253±1K ≈ $T^{**}$ [28], latter being the observed superheating temperature at different *T*-ramp-rates, and the former is the ideal thermodynamic (non-kinetic) transition temperature. $1/\chi$ (*H/M*) vs. *T* (fig.1a inset) displays ~ 2K hysteresis in (FC) cooling and (ZFC) warming magnetization data, along with their high-*T* Curie-fits down to supercooling ($T^*$) and superheating ($T^{**}$) [29] temperatures respectively. $T^*$ varies considerably with the cooling rate; obtained from the heat flow as 242K (5K/min) and 235K (15K/min). Thus, a *kinetic* (ramp-dependent) *T*-window characterizes the warming-cooling hysteresis; e.g., ($T^{**}$-$T^*$) ~ 18K from the heat-flow at 15K/min ramp-rate (fig.1b inset). The important observations confine the PM-FM phase-coexistence [30] entirely below the transition temperature $T_C$, above which the PMI phase exists alone. The precise delineation has bearing on the characterization of magneto-dielectric anomaly in our LCMO47. Supplementary data (fig.S1) shows magnetization (ZFC-warming and FC-cooling, both at 500 Oe) over the full temperature-range; the plateau at ~200K signifies the end (emergence) of PM-FM (FM-AFM) coexistence, while the rapid (FC/ZFC-split) high-level roll-off at lower *T*'s reflects predominantly FM ground state with small AFM admixture.

*3.2 Resistivity and Magneto-resistance*

Figure 2a (left-panel) shows the dc resistivity (at 0 and 5T fields) vs. temperature and fig.2b (right-panel) shows magneto-resistance $\rho(H)$ at key temperatures, taken on LCMO47. From the zero-field resistivity data, the temperature of its inflexion-point (highlighted as the larger/different symbol, denoting the metal-insulator (M-I) transition) is found to be same (to within 0.5K) as the $T_C$ obtained from the magnetization and thermal results. This is further confirmed by the gap-activated fit (~ $e^{\Delta/T}$) [31] to the high-temperature $\rho_0(T)$ right down to 254K; ruling out the emergence of any FMM phase above $T_C$. From fig.2b we observe that above ~1T field, the magneto-resistance (MR) below $T_C$ becomes nearly *T*-independent [19], along with a weaker *H*-dependence (especially for temperatures close to $T_C$). This indicates a change of the dominant mechanism responsible for the MR. We contend that the initial steeper MR($H$) is mainly caused by the PM-to-FM phase-conversion, while the latter/weaker change comes from the alignment of the (by then well-formed) FM-domains. This is consistent with the relative continuity of the milder MR($H$) behavior at lower temperatures (below 200K); here the PM to FM phase-conversion having mostly occurred at lower- or zero-field. Demise of the PM-FM phase-coexistence also registers as the local plateau in $M(T)$ (fig.S1) and the second inflexion point in MR($T$), both at ~200K (fig.S2). Further increase of MR($T$) (fig.S2) below ~125K and a net observable rise in $\rho$(0T) below ~30K respectively signify the now-dominant contribution of the small CO-AFM-insulating phase, and the growth of its contribution to resistivity.

*3.3 Permittivity and Magneto-Dielectricity*

Figure 3 shows the real permittivity of LCMO47 over 200-300K at benchmark frequencies across the 30Hz-1MHz measurement range. High values and strong frequency dependence at low frequencies signify dominantly extrinsic/inter-grain regime, delineating it from the high-frequency-- essentially intrinsic/intra-grain response, c.f., [32-34] as discussed in the next section. Inset shows the dielectric data at 100 kHz under zero- and 5T fields; the peak near 250K (corresponding to the zero-field M-I transition) does not indicate a ferroelectric (FE) transition; the post-peak part [[illegible] $(T\text{-}T_C)^{-1}$] is of "non-Curie" character and remains weakly frequency-dependent. The anomaly observed here has no *polarization* counterpart [35], yet qualitatively it is a more robust signature of magneto-dielectricity than the magneto-dielectric $\Delta\varepsilon'$-step signature (typically $\leq$ 15%) reported in "polar" multiferroics (c.f., $BiMnO_3$ and $YMnO_3$ [36]). The in-field (5T) permittivity exhibits a wide hump at 275K (the temperature where $\rho$(5T) has the inflexion point, denoting the in-field M-I transition). Besides this $T$-correspondence, intrinsic effects of magnetic-phase coexistence and their interface are qualitatively different on $\varepsilon'_{0T}(T)$ and $\rho_{0T}(T)$; affecting former directly via boosting local-insulator behavior and the latter through scattering of global charge-carriers and creation of percolative channels for them. The mixed-state of insulating (PM) and metallic (FM) regions sustaining interfacial space-charges contributes to the capacitive-response via *intra-grain* Maxwell-Wagner (M-W) mechanism [32]. Here, *local* isostructural variations (across the FM-PM interfacial-regions) too boost the polarizability ($\varepsilon'_{0T}$), also without realizing a net *bulk* polarization or altering the percolative/conducting paths. Acting as extended defects though, both these interface-effects provide increased scattering of the charge-carriers, adding to the dc resistivity $\rho_{0T}$ [37]. Consequently, field-dependent topological rearrangement of coexistent phases (FM tubules in PM matrix say, [14]) and of the associated local isostructural variations across their "domain-walls" reflect in the (ac) permittivity *directly*, and alter the (dc) resistance mainly via their across-the-sample percolative-reorganization; manifesting $MR_{dc}$-independent (magneto-dielectric)$_{ac}$ effect at high-frequencies. An across-the-specimen uniform-distortion of the $Mn\text{-}O_6$ octahedra ought to produce a net electrical polarization [22, 38] and bulk ferroelectricity, not yet observed (not in the present work either; magnetic phase-coexistence forbidding the "uniformity" requirement). Thus, the (intrinsic) magnetic phase-disorder here is responsible both for the observed magneto-dielectricity (albeit *intra-granular* "M-W" kind) and for the lack of long-range electrical order [35].

A neat segregation of extrinsic (inter-grain) and intrinsic (mainly FM-PM) contributions is illustrated by the dielectric response spectra at key temperatures (fig.4a, zero-field) and at 252K (fig.4b, under 0 & 5T fields). In fig.4a, from the spectrum at 252K (peak temperature of $\varepsilon'_{0T}(T)$, fig.3), we notice sharply-distinct extrinsic (low-$f$) and intrinsic (high-$f$) regimes, which appear as merged-over in the 300K spectrum. Here, extrapolating the low-frequency behavior ($\varepsilon_{ext}' \sim \omega^{-\alpha}$) up to 100 kHz (say) allows estimation of intrinsic to extrinsic response-ratio ($\varepsilon'/\varepsilon'_{ext}$) at benchmark $T$ and $H$. This zero-field metric near $T_C$ ($\sim 35000|_{252K}$) hugely outweighs that at room temperature ($\sim 200|_{300K}$) and much below $T_C$ ($\sim 3000|_{200K}$), as well as that under the field near $T_C$ ($\sim 6700|_{252K,5T}$, fig.4b); underlining a dominant contribution to the intrinsic response from the FM-PM coexistence. Imaginary permittivity is found (not explicitly shown here) to have the characteristic Maxwell-Wagner behavior ($\varepsilon'' \sim \omega^{-1}$) [32-33] over the full spectral-range. Cole-Cole plots ($\varepsilon''$ vs. $\varepsilon'$, fig.4a inset) delineate two power-law regimes; both distinctly formed at 200K and 252K, and overlapped at 300K. They signify low-frequency (inter-grain) and high-frequency (intra-grain) M-W conductive behaviors sans relaxations (absence of semicircles). Peaks in the loss-tangent $\tan\delta = \varepsilon''/\varepsilon'$ at 252K under zero- and 5T field (fig.4b, right $y$-axis) mark switchovers between extrinsic/intrinsic responses, and *not the relaxation* of either. Large values of the conductive losses here vs. the normal dielectric losses are not unexpected [17, 39]. From fig.4b inset, note that the low-frequency (extrinsic) regime exhibits comparable magnitudes (~100%) of magneto-capacitance [MC=($\varepsilon'_{5T}/\varepsilon'_{0T}$ -1)] and magneto-losses [ML=($\tan\delta_{5T}/\tan\delta_{0T}$ -1)], whereas the intrinsic (> ~3kHz) regime features a broadband MC ≈ -41% >> ML ≈ 2.7%. These results provide clear evidence that high-frequency (intrinsic) magneto-dielectricity and dc magneto-resistivity are uncoupled below $T_C$.

*3.4 Magneto-capacitance versus Magneto-resistance*

We now make the above arguments more precise, as regards the effects of magnetic phase-disorder on the (dc) resistive and (ac) capacitive responses in LCMO47, revealed by the application of the external magnetic field. As noted above, the 5T field almost completely removes the PM-FM phase coexistence, and provides baseline $T$-dependences in permittivity and resistivity, reflecting only the uniform isostructural changes across the PMI-FMM metal-insulator transition. Thus, by extracting the full dynamic range of the phase-coexistence-contribution, dc magneto-resistance [MR=($\rho_{5T}/\rho_{0T}$ -1)] and 100 kHz magneto-capacitance [MC=($\varepsilon'_{5T}/\varepsilon'_{0T}$-1)] together isolate the trivial (MR-determined MC) regime (> $T_C$). To this end, we show in fig.5 these evaluated parameters, with features otherwise buried under large baselines in the raw data (figs.2-3). Remarkable hand-in-hand tracking of the two (MR and MC) in the PMI state above ~250K contrasted with their independent evolution below cannot be overemphasized. While the MC shows a sharp & large (-41%) maximum at 252K, and a smaller one at 214K, MR saturates to a broad & high plateau (-50%) by 225K. Huang *et al.* [19] have reported via neutron diffraction the AFM ordering at 230K, with a different crystallographic phase in this same composition. These MC-maxima signify climaxes of the magnetic phase-coexistence-induced electrical-inhomogeneity effects. Demarcation of distinct temperature regimes is further facilitated by the plot of MC vs. MR in the inset. Here, the PMI state is distinguished by the straight-line fit to the high-temperature (> $T_C$) MC vs. MR, denoting trivial MR-driven magneto-capacitance. The sub-$T_C$ "parameter-space" clearly indicates ac magneto-capacitance (MC) not dictated by the dc magneto-resistance (MR); intra-grain FM-PM contribution to the dielectric response (below $T_C$, differently field-dependent) does not have an equivalent/one-to-one correspondence in dc resistivity.

## Conclusions

In conclusion, thermodynamic and magneto- transport/dielectric investigations of LCMO47 characterize its magneto-isostructural phase transition at ferromagnetic $T_C$ =253K precisely as disorder-broadened first-order type, with para-ferromagnetic phase-coexistence only below $T_C$. Consequent topographic phase- & structural- inhomogeneity host intra-granular Maxwell-Wagner effect. This manifests in intrinsic (< $T_C$) high-frequency (> ~kHz) magneto-dielectricity; with comparably little magneto-losses, and uncoupled to (dc) magneto-resistance over the same temperature-field regime. Our study also suggests the saturation of real (i.e., MR-independent) magneto-dielectricity with near-disappearance of FM-PM phase-coexistence beyond ~$O$(T) fields. Therefore, within the specified ($T$, $H$, $\omega$)-domain, well-known magnetic phase-disorder hosts the non-polar yet appreciable magneto-dielectricity in this magneto-resistive system, amenable to enhancements by materials engineering.

## Acknowledgements

We thank Alok Banerjee, Devendra Kumar, and Kranti Kumar for the magnetization data & tips. Rajeev Rawat and Sachin Kumar are thankfully acknowledged for magneto-resistance discussion & measurements. Dr. P. Chaddah is acknowledged for his scientific support and encouragement.

## Figure Captions

**1.** Magnetization $M(T)$ (top panel) and heat-capacity $C_p(T)$ (bottom panel) of LCMO47 across $T_C$ ~ 253K. Inset (a) shows Curie-fits ($1/\chi \sim T$) in warming and cooling cycles. Inset (b) depicts the indifference to the $T$-ramp rate, of the superheating temperature $T^{**} \approx T_C$, and strong kinetics of the supercooling temperature $T^*$ (numbers on the thermograms denote ramp-rates in K/min).

**2.** Resistivity of LCMO47 under zero and 5T fields (left panel). Inflexion point in $\rho_{0T}(T)$ at $T_C$ marks the insulator to metal transition, above which the gap-activated (~ $e^{\Delta/T}$) dependence fits exactly. Right panel-- Magneto-resistance MR($H$) with weaker field-dependence and relative $T$-invariance below $T_C$ and above ~1T fields possibly reflects the regime of FM-domains' alignment.

**3.** Temperature-dependence of zero-field real dielectric constant for LCMO47 at benchmark frequencies. At high frequencies, anomalies near the FM-$T_C$ and ~215K are both non-polar/non-relaxor type. Inset compares $\varepsilon'(T)$ at 100 kHz under zero and 5T fields, latter apparently devoid of the FM-PM and AFM-FM phase-coexistence contributions.

**4.** (a) Zero-field dielectric spectra at 200K ($\ll T_C$), 252K ($\sim T_C$), and 300K ($\gg T_C$). Extrinsic (inter-granular Maxwell-Wagner) response is evident at low (sub-kHz) frequencies, and in the Cole-Cole plot (inset). (b) Dielectric spectra at 252K under 0 and 5T fields (left $y$-axis). Peaks in tan$\delta$ (right $y$-axis) signify extrinsic/intrinsic switchover. Inset reflects qualitatively different field-effects on the capacitance and losses in the extrinsic ($|MC| \sim |ML|$) and intrinsic ($|MC| \gg |ML|$) regimes.

**5.** Magneto-resistance (MR) [$\rho_{5T}/\rho_{0T}$ -1] and magneto-capacitance (MC) [$\varepsilon'_{5T}/\varepsilon'_{0T}$ -1] under 5T field clearly delineate the pure PM-insulator phase from the FMM-PMI phase-coexistence below $T_C$. Inset (MC vs. MR) "phase-diagram" precisely extracts the coexistence-window, featuring ≈15% variation in magneto-dielectricity, impervious to the magneto-resistance (constant to within ~3%).

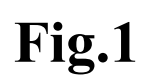

**Fig.2**

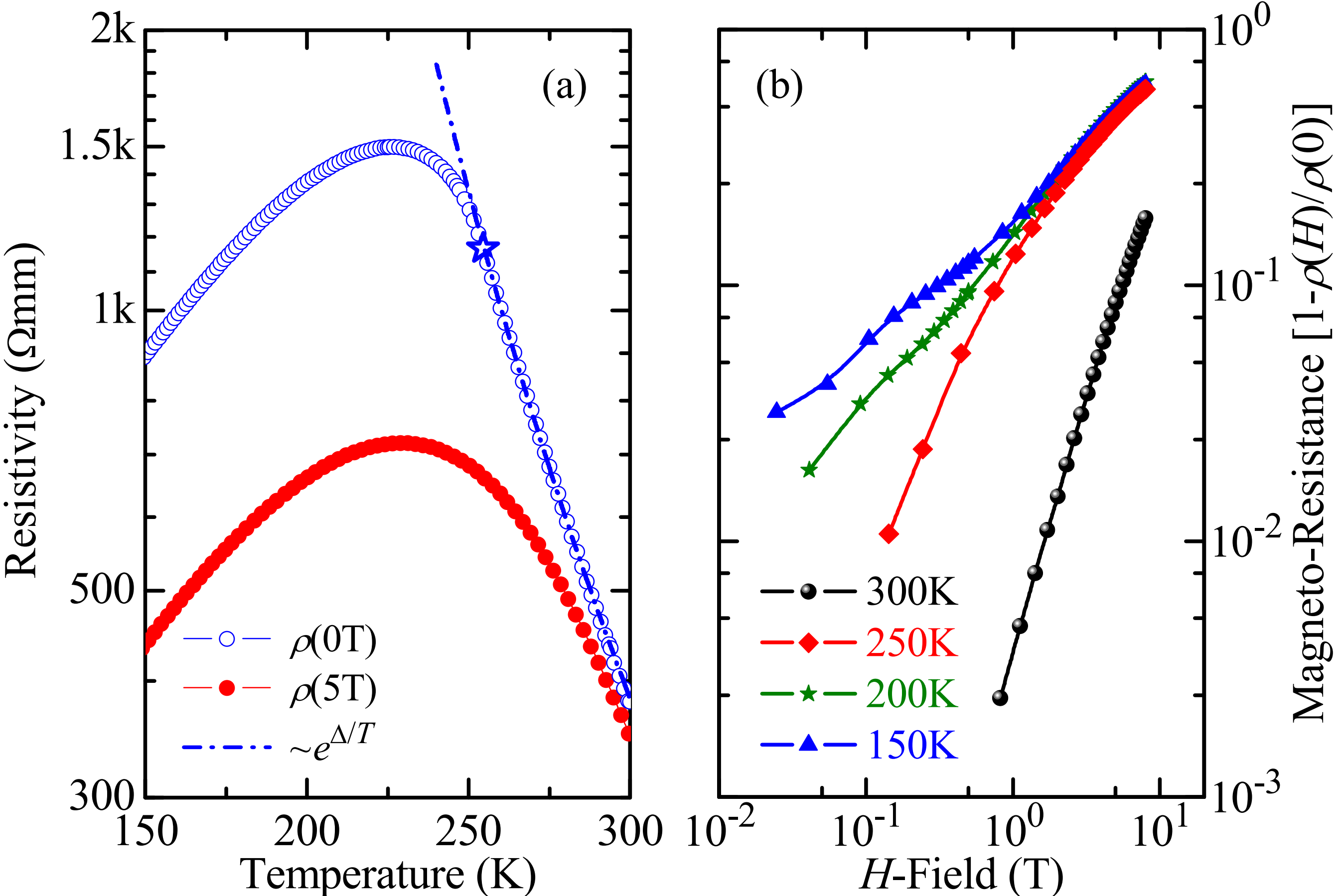

**Fig.3**

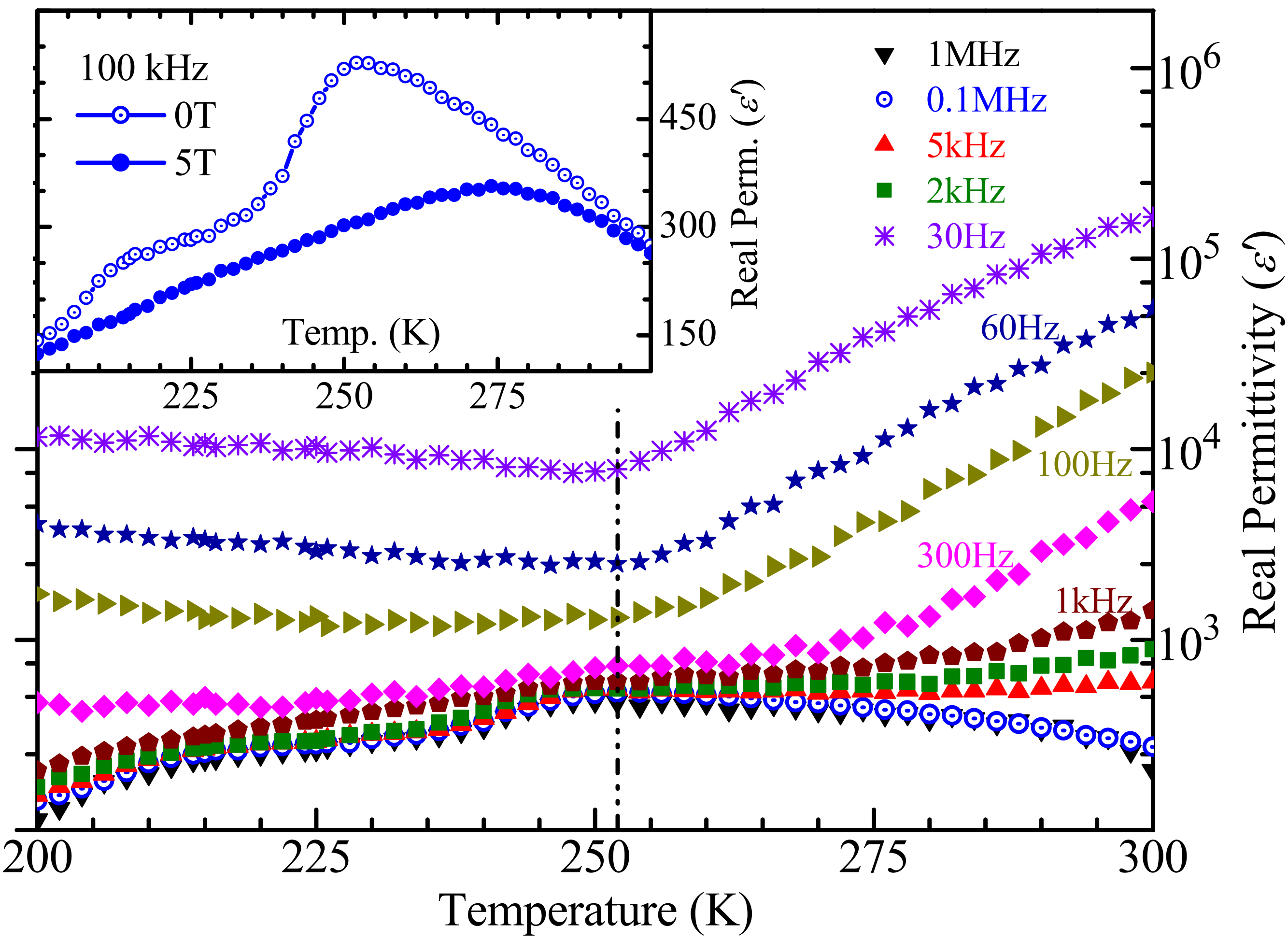

**Fig.4**

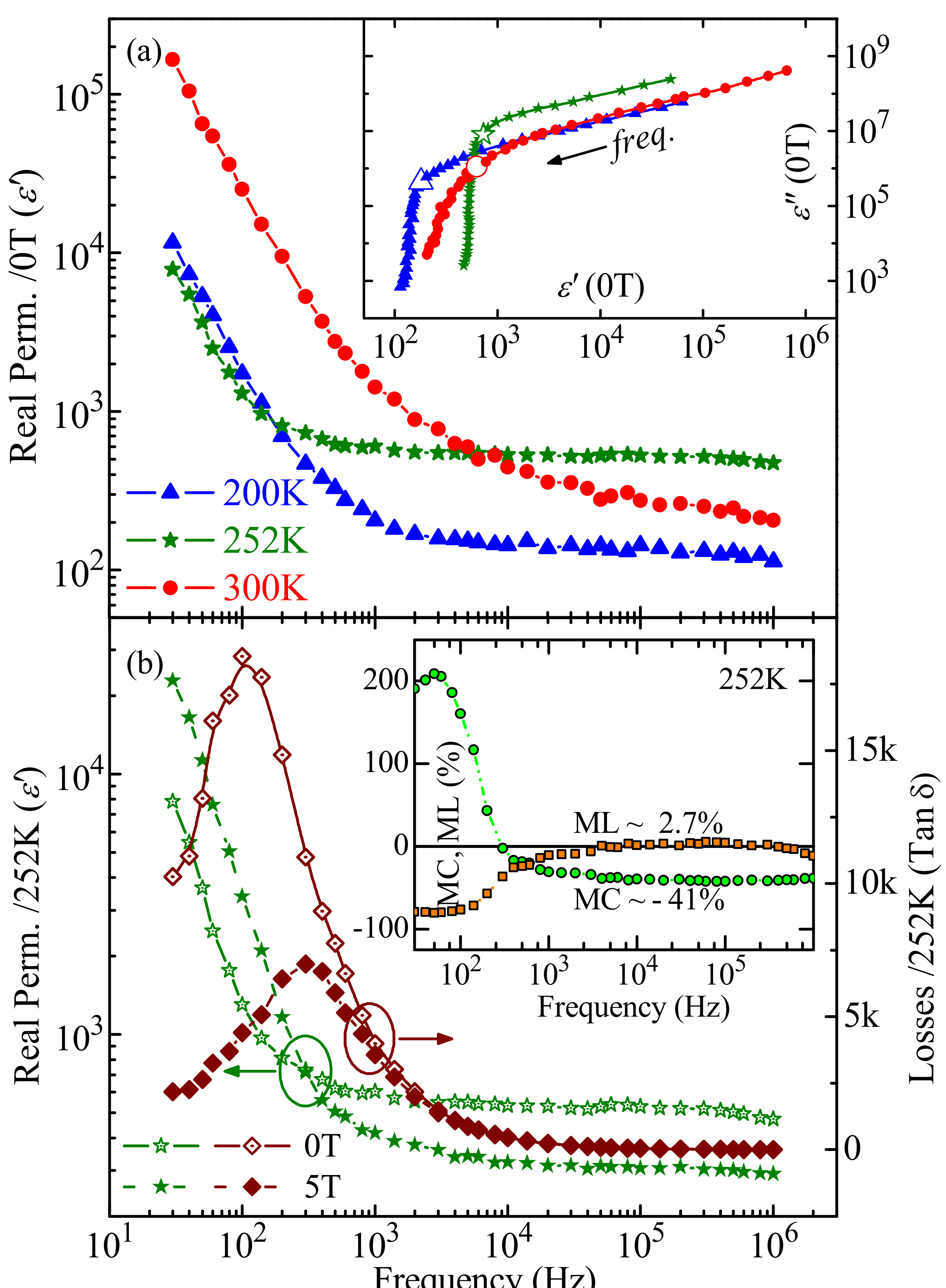

**Fig.5**

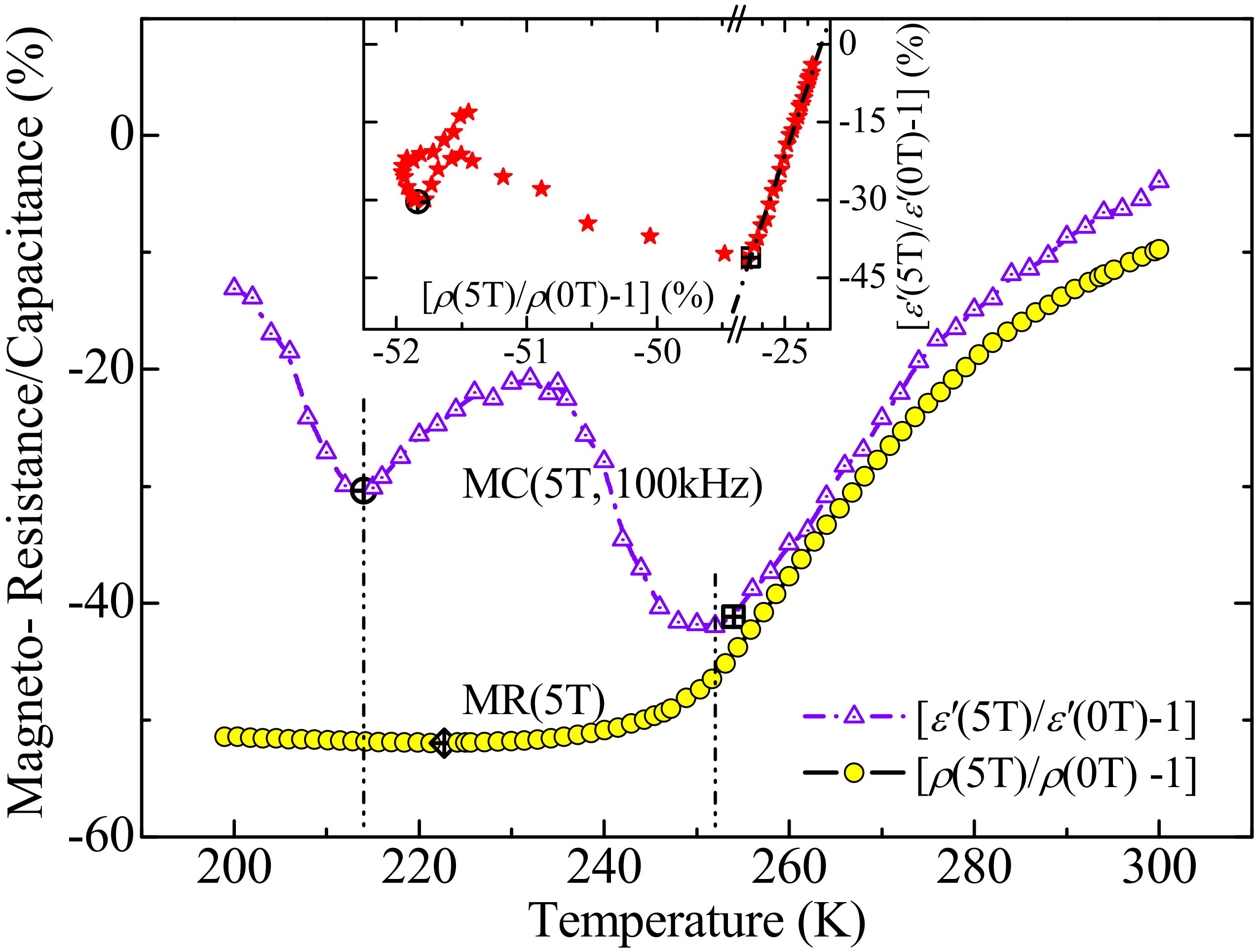

**Supplementary Figures**

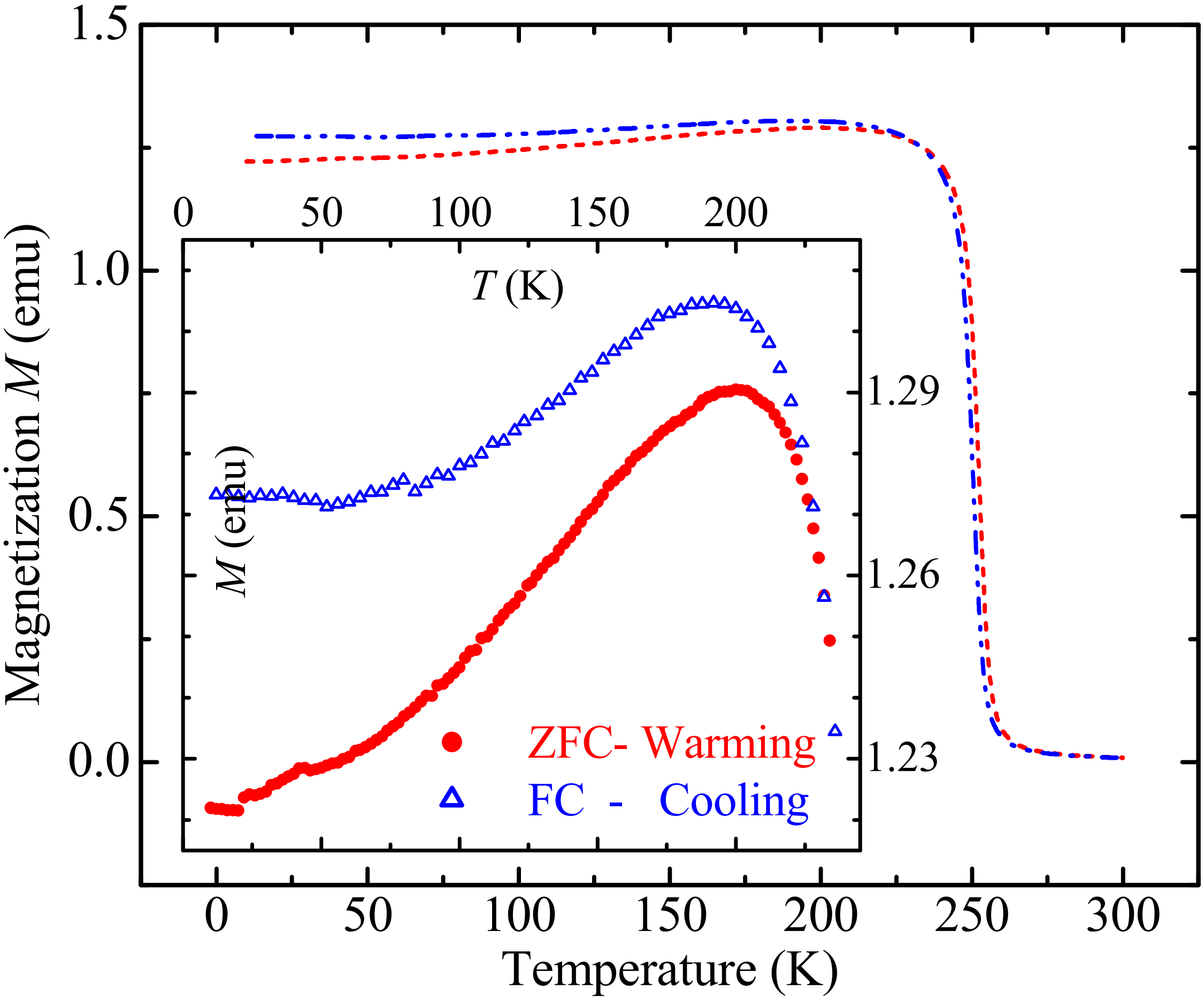

**Fig.S1.** Magnetization data (ZFC-warming and FC-cooling, both at 500 Oe field) over the full measurement temperature range, with zoomed-in low-$T$ data in the inset. Here, the predominantly ferromagnetic character of the ground state, with an admixture of the CO-AFM phase is evident from the small yet detectable drop below ~200K in $M(T)$, finally rolling-off to a high-value at the lowest temperatures.

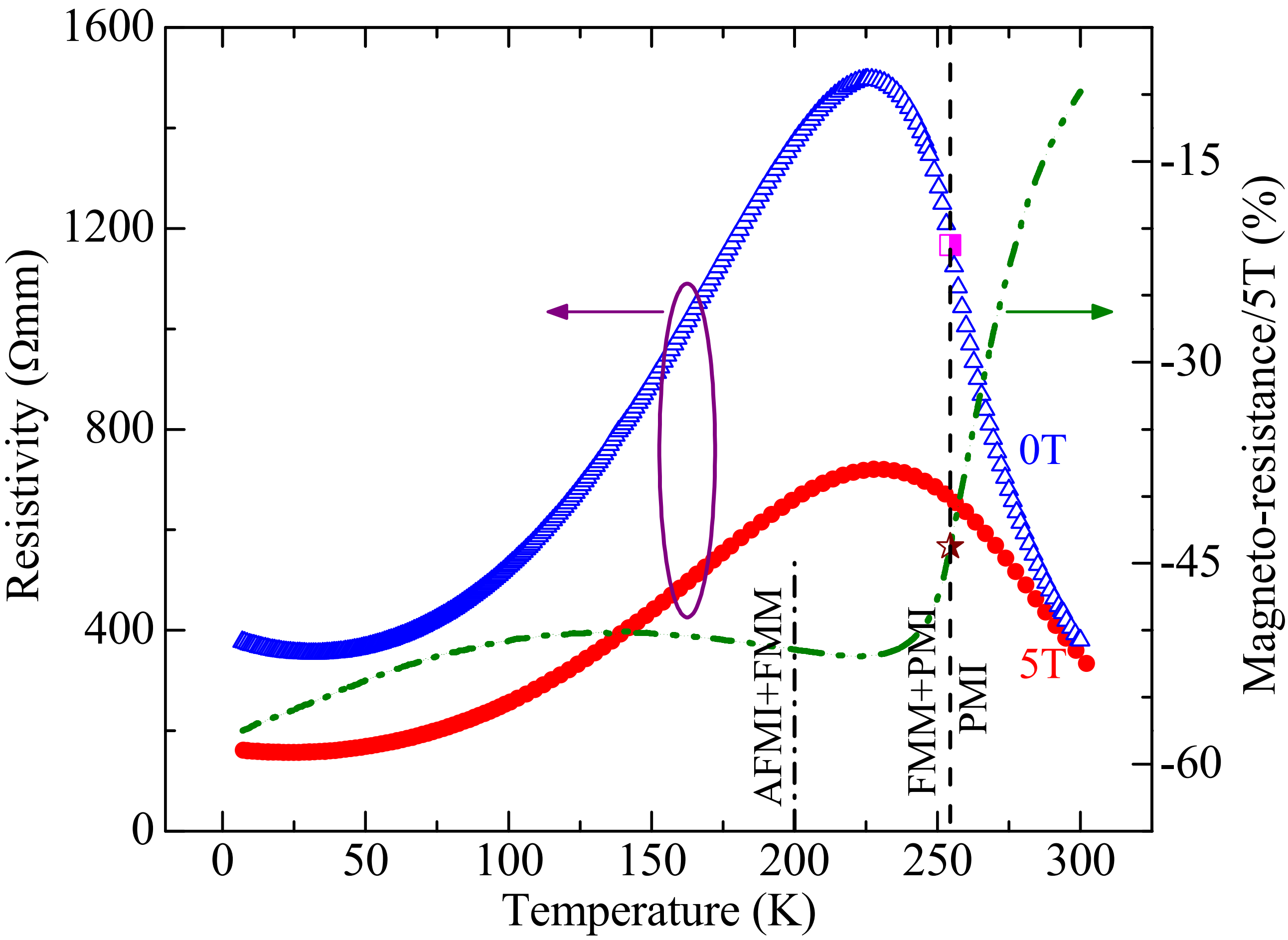

**Fig.S2.** Resistivity and magneto-resistance shown over the full measurement-temperature range. A net observable rise in $\rho$(0T) below ~30K and the "post-plateau" increase in the MR below ~125K mark the detectable insulating-contribution from the small CO-AFM phase, whose emergence is witnessed below ~200K in supplementary fig.S1.